\documentclass[12pt]{article}
\pdfoutput=1
\usepackage{amsmath}
\usepackage{epsfig}
\newcommand{\be}{\begin{equation}}
\newcommand{\ee}{\end{equation}}

\newcommand{\no}{\noindent}
\newcommand{\ce}{\begin{center}}
\newcommand{\nc}{\end{center}}

\makeatletter

\newcommand{\Rmnum}[1]{\expandafter\@slowromancap\romannumeral #1@}
\makeatother

\makeatletter
\@addtoreset{equation}{section}
\makeatother
\renewcommand{\theequation}{\thesection.\arabic{equation}}

\def\sqr#1#2{{\vcenter{\vbox{\hrule height.#2pt
 \hbox{\vrule width.#2pt height#1pt \kern#1pt
 \vrule width.#2pt} \hrule height.#2pt}}}}

\def\operp{\hbox{${\kern+.25em{\bigcirc}
\kern-.85em\bot\kern+.85em\kern-.25em}$}}

\def\lsim{\;\raise0.3ex\hbox{$<$\kern-0.75em\raise-1.1ex\hbox{$\sim$}}\;}
\def\gsim{\;\raise0.3ex\hbox{$>$\kern-0.75em\raise-1.1ex\hbox{$\sim$}}\;}
\def\no{\noindent}

\def\ce{\centerline}
\def\ve{\vfill\eject}
\def\rdots{\mathinner{\mkern1mu\raise1pt\vbox{\kern7pt\hbox{.}}\mkern2mu
 \raise4pt\hbox{.}\mkern2mu\raise7pt\hbox{.}\mkern1mu}}

\usepackage{indentfirst}

\def\e e{$e^+ e^-$ }

\newcommand{\ket}[1]{\left| #1 \right>} 
\newcommand{\bra}[1]{\left< #1 \right|} 
 
\usepackage{slashbox}
\usepackage{pict2e}

\usepackage{slashed}
 
 \usepackage{mathrsfs}
 
\def\subequations{\refstepcounter{equation}%
\edef\@savedequation{\the\c@equation}%
\@stequation=\expandafter{\theequation}
\edef\@savedtheequation{\the\@stequation}
\edef\oldtheequation{\theequation}%
\setcounter{equation}{0}%
\def\theequation{\oldtheequation\alph{equation}}}%

\def\endsubequations{%
\setcounter{equation}{\@savedequation}%
\@stequation=\expandafter{\@savedtheequation}%
\edef\theequation{\the\@stequation}\global\@ignoretrue}

\DeclareMathAlphabet{\mathpzc}{OT1}{pzc}{m}{it}

\begin{document}

\ce{\bf The Preon Sector of the SLq(2) (Knot) Model}
\vskip.3cm

\ce{\it Robert J. Finkelstein}
\vskip.3cm

\ce{Department of Physics and Astronomy}
\ce{University of California, Los Angeles, CA 90095-1547}

\vskip1.0cm 

\no {\bf Abstract.} We describe a Lagrangian defining the preon sector of the knot model. The preons are the elements of the fundamental representation of SLq(2), and unexpectedly agree with the preons conjectured by Harari and by Shupe. The leptons, neutrinos, up and down quarks, described as $j=3/2$ representations, and the electroweak vectors, described as $j=3$ representations, of SLq(2) also have the preon composition required by the schemes of Harari and of Shupe. The coupling constants and masses required by the preon Lagrangian introduce form factors that are simple functions of the knot model parameters. As previously shown the knot model has similarly provided a possible parametrization of the masses and electroweak coupling constants (Kobayashi-Maskawa matrix) of the standard model. There is an alternative formulation of the kinematics permitting possible additional isotopic partners of the quarks and the neutrinos.


\thispagestyle{empty}
\setcounter{page}{0}

\newpage

\section{Introduction}

In the knot model, here understood as a SLq(2) extension of the standard model, the elementary fermions with isotopic spin $t=1/2$ are quantum trefoils lying in the $j=3/2$ representation of SLq(2). The electroweak vectors with $t=1$ belong to the $j=3$ representation. The adjoint and fundamental representations define hypothetical particles that may be described as preons. Some of the possible experimental evidence of the $j=3/2$ and $j=3$ representations has been discussed.$^{(1)}$ Here we would like to present a corresponding discussion of the preon sector of the model, described by the $j=1/2$ fundamental representation of fermionic preons and by the $j=1$ adjoint representation of bosonic vectors.

This study began as an effort to examine the compatibility of the SLq(2) symmetry with the standard model. In physical terms this compatibility would mean that the elementary particles are in some sense knotted, since SLq(2) describes the knot algebra. It turns out there is indeed empirical support for a description in which the simplest particles are the simplest knots and in which they are described by the $j=3/2$ representation of SLq(2). When one then looks at the adjoint ($j=1$) and fundamental $(j=1/2)$ representations, one unexpectedly finds exactly the preon structure proposed earlier by Harari and by Schupe. The Harari, Shupe, and knot models are all based on the same experimental data, but Harari and Schupe aim to produce the simplest possible basis for a more fundamental theory, while the goal in the third approach is to construct a SLq(2) extension of the standard model. It is remarkable that these three different approaches lead to the same preon model.

In the SLq(2) model, the creation operators of all the composite particles, described by representations D$_{mn}^j$ of SLq(2), are polynomials in the creation operators, D$_{mn}^{^{1/2}}$\hspace{0.1cm}, $(a,b,c,d)$, of the four preons, which in turn are the elements of the fundamental representation with charges $-\frac{e}{3}(m+n)$. In the particular case of leptons, neutrinos, down and up quarks, the creation operators, D$_{mn}^{^{3/2}}$\hspace{0.1cm}, are monomials proportional to $a^3$, $c^3$, $ab^2$, $cd^2$, respectively. There is then one charged preon, $a$, with charge $-\frac{e}{3}$ and its antiparticle $d$ while there is one neutral preon, $b$, and its antiparticle $c$, just as previously proposed by Harari and by Shupe. The $a^3,c^3,ab^2,cd^2$ describe the creation operators and preon content of the four classes of elementary fermions, also precisely as proposed by Harari and by Shupe.

The preon operators $(a,b,c,d)$ may be interpreted as creation operators for either particles or twisted flux loops with one crossing, and $(a^3,c^3,ab^2,cd^2)$ may be interpreted as creation operators for three preons or for a trefoil flux loop with three crossings. The two complementary descriptions may be reconciled as describing a composite particle composed of three preons bound by a trefoil flux tube.

\section{The Elementary Fermions and Electroweak Vectors in the Knot Model}

To obtain the SLq(2) extension of the standard model, replace the field operators, $\Psi$, of the standard model by
\begin{equation}
\hat{\Psi}_{mp}^j=\Psi \mbox{D}_{mp}^j(\Psi)
\end{equation}
where the $\Psi$ may be an elementary fermion, weak vector, or Higgs field operator. The $\mbox{D}_{mp}^j$ are elements of the irreducible representations of the knot algebra SLq(2). The normal modes of $\hat{\Psi}_{mp}^j$ define the field quanta of the extended model and these field quanta will be called ``quantum knots."

We postulate a correspondence between quantum knots and oriented classical knots according to
\begin{equation}
\left( j, m, p \right) = \frac{1}{2}\left(N, w, r+o \right)
\end{equation}
where $\left(N, w, r\right)$ are (the number of crossings, the writhe, and the rotation, respectively) of the 2d-projection of an oriented classical knot. Since the $\left( N , w, r \right)$ are integers, the factor $1/2$ is needed to allow half-integer representations of SLq(2).  Since $2m$ and $2p$ are of the same parity, while $w$ and $r$ are topologically constrained to be of opposite parity, $o$ is an odd integer that we set $=1$ for a quantum trefoil.$^{(1)}$

Eq. (2.2) restricts the states of the quantum knot to only those states of the full $2j+1$ dimensional representations that correspond to the 2d-spectrum $(N, w, r)$ of a corresponding classical knot. The interesting consequences of not postulating (2.2) will be considered later.

The defining SLq(2) algebra is invariant under gauge transformations U$_a(1)\times$U$_b(1)$ that induce the following gauge transformations on the D$_{mp}^j$$^{(1)(2)}$
\begin{displaymath}
\mbox{D}_{mp}^{j \hskip0.1cm \prime} = e^{i m \varphi_m}e^{ip \varphi_p} \mbox{D}_{mp}^j \tag{2.3a}
\end{displaymath}
\begin{displaymath}
\hspace{1.68cm}= \mbox{U}_m(1)\times \mbox{U}_p(1) \mbox{D}_{mp}^j \tag{2.3b}
\end{displaymath}
and hence on the $\hat{\Psi}_{mp}^j$:
\begin{equation}
\setcounter{equation}{4}
\hat{\Psi}_{mp}^{j \hskip0.1cm \prime} = \mbox{U}_m(1)\times \mbox{U}_p(1) \hat{\Psi}_{mp}^j
\end{equation}
For physical consistency the field action is required to be invariant under Eq. (2.4) since the U$_{a}(1)\times$U$_b(1)$ transformations do not change the defining algebra.$^{(2)}$

Then there will be the following Noether charges that may be described by (2.2) as writhe and rotation charges
\begin{displaymath}
Q_w \equiv - k_w m = -k_w \frac{w}{2} \tag{2.5a}
\end{displaymath}
\begin{displaymath}
\hspace{0.3cm} Q_r \equiv -k_r p = -k_r \frac{r+1}{2} \tag{2.5b}
\end{displaymath}

\setcounter{equation}{5}

We assume that $k=k_w=k_r$ is a universal constant with the dimensions of an electric charge and with the same value for all trefoils.

The knot picture is more attractive if the simplest particles are the simplest knots.
We therefore consider the possibility that the most elementary fermions with isotopic spin $t=1/2$ are the most elementary quantum knots, the quantum trefoils with $N=3$. This possibility is supported by the following empirical observation
\begin{equation}
(t, -t_3, -t_0) = \frac{1}{6}(N, w, r+1)
\end{equation}
which is satisfied by the four classes of elementary fermions described by $(1/2, t_3, t_0)$ and the four quantum trefoils described by $(3, w, r)$ and shown by the row to row correspondence in Table (2.1).
\newpage
\begin{center}
\begin{tabular}[h]{ccccc|ccccc}
\multicolumn{5}{c}{\underline{Standard Model}} &
\multicolumn{5}{c}{\underline{Quantum Trefoil Model}} \\
\underline{$(f_1,f_2,f_3)$} & \underline{$t$} &
\underline{$t_3$} &
\underline{$t_0$} & \underline{$Q_e$} & \underline{$(w,r)$} &
\underline{$\mbox{D}^{N/2}_{\frac{w}{2}\frac{r+1}{2}}$} &
\underline{$Q_w$} & \underline{$Q_r$} & \underline{$Q_w+Q_r$} \\
$(e,\mu,\tau)_L$ & $\frac{1}{2}$ & $-\frac{1}{2}$ & $-\frac{1}{2}$
& $-e$ & (3,2) & $\mbox{D}^{^{3/2}}_{\frac{3}{2}\frac{3}{2}}$ &
$-k\left(\frac{3}{2}\right)$ & $-k\left(\frac{3}{2}\right)$ &
$-3k$ \\
$(\nu_e,\nu_\mu,\nu_\tau)_L$ & $\frac{1}{2}$ & $\frac{1}{2}$ &
$-\frac{1}{2}$ & 0 & (-3,2) & $\mbox{D}^{^{3/2}}_{-\frac{3}{2}
\frac{3}{2}}$ & $-k\left(-\frac{3}{2}\right)$ &
$-k\left(\frac{3}{2}\right)$ & 0 \\
$(d,s,b)_L$ & $\frac{1}{2}$ & $-\frac{1}{2}$ & $\frac{1}{6}$ &
$-\frac{1}{3}e$ & (3,-2) & $\mbox{D}^{^{3/2}}_{\frac{3}{2}-\frac
{1}{2}}$ & $-k\left(\frac{3}{2}\right)$ & $-k\left(-\frac{1}{2}
\right)$ & $-k$ \\
$(u,c,t)_L$ & $\frac{1}{2}$ & $\frac{1}{2}$ & $\frac{1}{6}$ &
$\frac{2}{3}e$ & (-3,-2) & $\mbox{D}^{^{3/2}}_{-\frac{3}{2}
-\frac{1}{2}}$ & $-k\left(-\frac{3}{2}\right)$ &
$-k\left(-\frac{1}{2}\right)$ & $2k$ \\  & & & & $Q_e =e(t_3+t_0)$ & & & $Q_w=-k\frac{w}{2}$ &
$Q_r = -k\frac{r+1}{2}$& \\
\end{tabular}
\end{center}
\begin{center}
\bf{Table 2.1}
\end{center}

We next require that the total charge of each quantum trefoil, $Q_w+Q_r$, agrees with the total charge of the corresponding fermion, $Q_e$. Then $k=\frac{e}{3}$.

There are four choices of $(N, w, r)$ corresponding to the four quantum trefoils and there are four choices of $(t, t_3, t_0)$ corresponding to the four classes of fermions, but there is only a single way to match the two sides of Eqn. (2.6). 

By (1.2) and (2.6) one may also write
\begin{equation}
(j,m,p) = 3(t, -t_3, -t_0)
\end{equation}
There are therefore two ways of reading the indices on D$_{mp}^j$, either topologically by (2.2) or empirically by (2.7).$^{(1)}$

To describe the electroweak vector bosons, note that the fermion-boson interaction acquires by (2.1) the form factor
\begin{equation}
\bar{\mbox{D}}_{m'' p''}^{3/2}\mbox{D}_{mp}^j \mbox{D}_{m'p'}^{3/2}
\end{equation}
where the initial and final factors describe fermions and the middle factor describes a vector boson. All three factors satisfy (2.3) and the product (2.8) is required to be invariant under U$_a(1)\times$U$_b(1)$ in order to preserve the invariance of the field action.

Then
\begin{displaymath}
m=m''-m' \tag{2.9$m$}
\end{displaymath}
\begin{displaymath}
p=p''-p' \tag{2.9$p$}
\end{displaymath}
and by (2.7)
\begin{displaymath}
m=-3(t_3''-t_3') \tag{2.10$m$}
\end{displaymath}
\begin{displaymath}
p=-3(t_0''-t_0') \tag{2.10$p$}
\end{displaymath}

We require that the invariance under the SU(2)$\times$U(1) satisfied by the standard model be preserved in the extended standard model. Then
\begin{displaymath}
t_3=t_3''-t_3' \tag{2.11$t_3$}
\end{displaymath}
\begin{displaymath}
t_0=t_0''-t_0' \tag{2.11$t_0$}
\end{displaymath}
The separate conservation of $t_3$ and $t_0$ also expresses the separate conservation of writhe and rotation charges demanded by U$_a(1)\times$U$_b(1)$ invariance. By (2.10) and (2.11)
\begin{equation}
\setcounter{equation}{12}
(m,p) = -3(t_3,t_0)
\end{equation}
We then find that (2.7) also holds for the electroweak vector bosons with $t=1$ and $j=3$:
\begin{equation}
(j,m,p) = 3 (t,-t_3, -t_0)
\end{equation}

\section{The Preon Representations$^{(1)}$}

The knot model describes the elementary fermions and electroweak vectors as quantum knots lying in the $j=3t$ representations. The fundamental $(j=1/2)$ and adjoint $(j=1)$ representations describe hypothetical preons and preonic vector bosons. If the four representations $j=(3, 3/2, 1, 1/2)$ are uniformly described by (2.2), then $j=N/2$, and $j=1/2$ implies $N=1$ crossings; and $j=1$ implies $N=2$ crossings. The adjoint and fundamental representations therefore do not correspond to knots but they do correspond to twisted loops that may be assigned writhe and rotation integers in the same way as for knots.

In all cases we shall assume that the first factor, $\Psi$, in the field operator $\Psi \mbox{D}_{mp}^j$ transforms under SU(2)$\times$U(1). For the adjoined knot factor we shall in all cases take the following explicit form of D$_{mp}^j$.$^{(2)}$
\begin{displaymath}
\mbox{D}^j_{mm'}=\sum_{\begin{array}{rllll} 0 & \le & s & \le &  n_+ \\ 0 & \le & t & \le & n_- \end{array}} \mbox{A}^j_{mm'}(q,s,t) \delta (s+t,n'_+) a^s b^{n_+-s} c^t d^{n_--t} \tag{3.1a}
\end{displaymath}
where
\begin{displaymath}
\mbox{A}_{mm'}^j \left(q,s,t \right) = \left[ \frac{\langle n_+^\prime \rangle_1 ! \langle n_-^\prime \rangle _1 !}{\langle n_+ \rangle_1! \langle n_- \rangle_1 !}\right]^{\frac{1}{2}} \frac{ \langle n_+ \rangle_1 !}{\langle n_+ - s \rangle_1 ! \langle s \rangle_1 !} \frac{\langle n_- \rangle_1 !}{\langle n_- - t \rangle_1 ! \langle t \rangle_1 !} \tag{3.1b}
\end{displaymath}
and
\begin{displaymath}
\langle n \rangle_1 = \frac{q_1^n -1}{q_1 -1} \hskip0.5cm \mbox{and} \hskip0.5cm \begin{array}{rcl} n_\pm & = & j \pm m \\ n_\pm^\prime & = & j \pm m^\prime \end{array} \tag{3.1c}
\end{displaymath}
Here the arguments satisfy the knot algebra:
\begin{equation}
\setcounter{equation}{2}
\begin{array}{rcl} 
ab & = & qba \\ ac & = & qca
\end{array}
\hspace{1.0cm}
\begin{array}{rcl}
bd & = & qdb \\ cd & = & qdc
\end{array}
\hspace{1.0cm}
\begin{array}{rcl}
ad-qbc & = & 1 \\ da-q_1cb & = & 1
\end{array}
\hspace{1.0cm}
\begin{array}{rcl}
bc & = & cb \\ q_1 & \equiv & q^{-1} 
\end{array}
\end{equation}
The $(a,b,c,d)$ are also elements of the preon representation:
\begin{equation}
\mbox{D}_{mm'}^{1/2} = \begin{pmatrix} a & b \\  c & d \end{pmatrix}
\end{equation}
Since the D$_{mm'}^j$ factor is in general a polynomial in the operators $(a,b,c,d)$, the knot field operator, with the adjoined D$_{mm'}^j$ factor, is itself in general a polynomial in $a,b,c,d$. We shall interpret $(a,b,c,d)$ as creation operators for the preons $(a,b,c,d)$ when operating on the vacuum or on higher states. If D$_{mm'}^j$ is a monomial, it will create a single state; otherwise it will create a superposition of several states.

If we denote the exponents of $a,b,c,d$ in (3.1a) by $(n_a,n_b,n_c,n_d)$ respectively, then $(n_a,n_b,n_c,n_d)$ are the numbers of $a, b, c, d$ preons in the corresponding state.

The sum in (3.1a) over $(s,t)$ may be rewritten as a sum over $(n_a,n_b,n_c,n_d)$ subject to
\begin{displaymath}
n_a + n_b + n_c + n_d = 2j \tag{3.4$j$}
\end{displaymath}
\begin{displaymath}
n_a + n_b - n_c -n_d = 2m \tag{3.4$m$}
\end{displaymath}
\begin{displaymath}
n_a - n_b + n_c - n_d = 2m' \tag{3.4$p$}
\end{displaymath}
But by (2.7) one has the empirical constraint
\begin{equation}
\setcounter{equation}{5}
(j,m,m') = 3 (t, -t_3,-t_0)
\end{equation}
and by (2.2) one has the topological constraint
\begin{equation}
(j,m,m') = \frac{1}{2}(N,w,r+o)
\end{equation}
We have seen how the relations (3.5) and (3.6) hold for the $j=3/2$ and $j=3$ representations. We now assume that they hold for all representations that we consider.
Then by (3.4) and (3.5) we have for the composite particle described by $(t,t_3,t_0)$
\begin{displaymath}
t = \frac{1}{6} (n_a + n_b + n_c + n_d) \tag{3.7$t$}
\end{displaymath}
\begin{displaymath}
t_3 = -\frac{1}{6} (n_a + n_b - n_c - n_d) \tag{3.7$t_3$}
\end{displaymath}
\begin{displaymath}
t_0 = -\frac{1}{6}(n_a - n_b + n_c - n_d) \tag{3.7$t_0$}
\end{displaymath}
and by (3.4) and (3.6) the same particle described by $(N,w,r+o)$
\begin{displaymath}
N= n_a + n_b + n_c + n_d \tag{3.8$N$}
\end{displaymath}
\begin{displaymath}
w = n_a - n_d + n_b - n_c \tag{3.8$w$}
\end{displaymath}
\begin{displaymath}
r+o = n_a - n_b + n_c - n_d \tag{3.8$r$}
\end{displaymath}

By (3.3) and (3.5) one also has the Table (3.1)
 \begin{center}
 \underline{\bf{Table 3.1}}
 \end{center}
\begin{minipage}[c]{0.45 \textwidth}
\hspace{4.0cm}
\begin{tabular}{c | c c c c}
 & $t$ & $t_3$ & $t_0$ & $Q$ \\
 \hline
 $a$ & $\frac{1}{6}$ & -$\frac{1}{6}$ & -$\frac{1}{6}$ & -$\frac{e}{3}$\\
 $b$ & $\frac{1}{6}$ & -$\frac{1}{6}$ & $\frac{1}{6}$ & 0  \\
 $d$ & $\frac{1}{6}$ & $\frac{1}{6}$ & $\frac{1}{6}$ & $\frac{e}{3}$\\
 $c$ & $\frac{1}{6}$ & $\frac{1}{6}$ & -$\frac{1}{6}$ & 0\\
 \end{tabular}
 \end{minipage}
 \hspace{2.0cm}
 \begin{minipage}[c]{0.45 \textwidth}
$Q = e(t_3+t_0)$
 \end{minipage}
 
 \vspace{0.75cm}
\noindent \underline{According to Table (3.1) there is one charged preon, $a$, with charge $-e/3$ and its anti-}

\noindent \underline{particle, $d$. There is one neutral preon, $b$, and its antiparticle, $c$}. These particles agree with the preons of the Harari-Shupe model.$^{(3)}$
 
 By Table (3.1) and equations (3.7) one then has for every term $(n_a,n_b,n_c,n_d)$ in (3.1a)
 \begin{displaymath}
 t= \sum_p n_pt_p \tag{3.9$t$}
 \end{displaymath}
 \begin{displaymath}
 t_3 = \sum_p n_p t_{3p} \tag{3.9$t_3$}
 \end{displaymath}
 \begin{displaymath}
 t_0=\sum_p n_p t_{0p} \tag{3.9$t_0$}
 \end{displaymath}
 \begin{displaymath}
 Q = \sum_p n_p Q_p \tag{3.9$Q$}
 \end{displaymath}
 where $p=(a,b,c,d)$.  Here $n_p$ is the number of $p$-preons and $(t_p,t_{3p}, t_{0p}, Q_p)$ are quantum numbers of the $p^{th}$ preon according to Table 3.1.
 
The composite particle described on the left side of (3.9) with quantum numbers $(t, t_3, t_0, Q)$ is a superposition of states, each state characterized by the same $(t, t_3, t_0, Q)$ but with varying numbers $(n_a,n_b,n_c,n_d)$ of $(a,b,c,d)$ preons. These relations are illustrated by the following tables both obtained directly from Eqn. (3.1) and describing the $(a,b,c,d)$ preons and the $(\ell,\nu,d,u)$ fermions.

\begin{center}
\bf{\underline{Preons ($j$ = $\frac{1}{2}$)}}
\end{center}
\begin{center}
\begin{tabular}{c | c c c c}
 & Q & $t_3$ &  $t_0$ & $\mbox{D}^{3t}_{-3t_3-3t_0}$ \\
 \hline
 a & -$\frac{e}{3}$ & -$\frac{1}{6}$ & -$\frac{1}{6}$ & $\mbox{D}^{^{1/2}}_{\frac{1}{2} \frac{1}{2}} \sim a$ \\
 b & 0 & -$\frac{1}{6}$ & $\frac{1}{6} $ & $\mbox{D}^{^{1/2}}_{\frac{1}{2} -\frac{1}{2}} \sim b$\\
 c & 0 & $\frac{1}{6}$ & -$\frac{1}{6}$ & $\mbox{D}^{^{1/2}}_{-\frac{1}{2} \frac{1}{2}} \sim c$\\
 d & $\frac{e}{3}$ & $\frac{1}{6}$ & $\frac{1}{6}$ & $\mbox{D}^{^{1/2}}_{-\frac{1}{2} -\frac{1}{2}} \sim d$\\
\end{tabular}
\end{center}
\begin{center}
\bf{Table 3.2}
\end{center}

\begin{center}
\bf{\underline{Fermions ($j$= $\frac{3}{2}$)}}
\end{center}
\begin{center}
\begin{tabular}{c | c c c c}
 & Q & $t_3$ & $t_0$ & $\mbox{D}^{3t}_{-3t_3-3t_0}$\\
 \hline
 $\ell$ & -$e$ & -$\frac{1}{2}$ & -$\frac{1}{2}$ & $\mbox{D}^{^{3/2}}_{\frac{3}{2} \frac{3}{2}} \sim a^3$\\
 $\nu$ & 0 & $\frac{1}{2}$ & -$\frac{1}{2}$ & $\mbox{D}^{^{3/2}}_{-\frac{3}{2} \frac{3}{2}} \sim c^3$\\
 d & -$\frac{1}{3}e$ & -$\frac{1}{2}$ & $\frac{1}{6}$ & $\mbox{D}^{^{3/2}}_{\frac{3}{2} -\frac{1}{2}} \sim ab^2$\\
 $u$ & $\frac{2}{3}e$ & $\frac{1}{2}$ & $\frac{1}{6}$ & $\mbox{D}^{^{3/2}}_{-\frac{3}{2}  -\frac{1}{2}} \sim cd^2$\\
 \end{tabular}
\end{center}
\begin{center}
\bf{Table 3.3}
\end{center}

Since the fermion knot factors D$_{mp}^{^{3/2}}$ are all monomials of the third degree, the corresponding composite particles contain three preons. \underline{The leptons and neutrinos are composed of} \underline{three $a$ and three $c$ preons, respectively. The down quarks are composed of one $a$ and two $b$} \underline{preons while the up quarks are composed of one $c$ and two $d$ preons}, again in agreement with Harari and Shupe.

\section{The Complementary Models$^{(1)}$}

Since $(N,w,r+o) = 2(j,m,m')$, the equations (3.4) may also be read as knot relations (3.8). As there are only three equations to determine the four $(n_a,n_b,n_c,n_d)$, the composite particle, described by either $(t,t_3,t_0)$ or $(N,w,r)$, is in general a superposition of several components with different sets of $(n_a,n_b,n_c,n_d)$.

Equations $(3.8N)$ states that the total number of preons equals the number of crossings $(N)$. Since we assume that the preons are fermions, the knot describes a fermion or a boson depending on whether the number of crossings is odd or even.

Since $a$ and $d$ are antiparticles with opposite charge and hypercharge, while $b$ and $c$ are neutral antiparticles with opposite values of the hypercharge, we may introduce the preon numbers
\begin{equation}
\nu_a = n_a - n_d
\end{equation}
\begin{equation}
\nu_b = n_b - n_c
\end{equation}
Then $(3.8w)$ and $(3.8r)$ may be rewritten as
\begin{eqnarray}
\nu_a + \nu_b &=& w \hskip0.3cm \left( = -6 t_3 \right) \\
\nu_a - \nu_b &=& r+o \hskip0.3cm \left(= -6t_0 \right)
\end{eqnarray}

By (4.3) and (4.4) the conservation of the preon numbers and of charge and hypercharge is equivalent to the conservation of the writhe and rotation, which are topologically conserved at the classical level. In this respect, these quantum conservation laws correspond to the classical conservation laws.

The SLq(2) equations (3.8) hold for all representations and therefore for preons as well as for knots, although the preons are twisted loops rather than knots. If the indices $(N, w, r)$ for the fermionic preons are determined in the same way as for knots, one finds $N=1$, $w=\pm1$, and $r=0$. Then by (3.8$r$) the odd integer, $o$, is for preons.
\begin{equation}
o = n_a + n_b - n_c -n_d
\end{equation}
It follows that $o=1$ for $a$ and $b$, and that $o=-1$ for the anti-particles, $d$ and $c$.

Viewed as a knot, a fermion becomes a boson when the number of crossings is changed by attaching or removing a curl. This picture is consistent with the view of a curl as an opened preon loop, and of the elements of the knot algebra $(a,b,c,d)$ as creation operators of preon loops.

Corresponding to the representations of the four elementary fermions as three preon states, there is the complementary representation of the four trefoils as composed of three overlapping preon loops as shown in Fig. 4.1. In interpreting Fig. 4.1 note that the two lobes of all the preons make opposite contributions to the rotation, $r$, so that the total rotation of each preon vanishes. When the three $a$-preons and $c$-preons are combined to form leptons and neutrinos, respectively, each of the three labelled circuits is counterclockwise and contributes $+1$ to the rotation while the single unlabeled shared circuit is clockwise and contributes $-1$ to the rotation so that the total $r$ for both leptons and neutrinos is $+2$. For the quarks the three labelled loops contribute $-1$ and the shared loop $+1$ so that $r=-2$.

Written in terms of $(N,w,r)$ and $(N,w,r)_p$ the equations describing the composite particles are
\begin{eqnarray}
N=\sum_p n_p N_p \nonumber \\
w = \sum_p n_p w_p \\
\tilde{r} = \sum_p n_p \tilde{r}_p \nonumber
\end{eqnarray}
where $p=(a,b,c,d)$ and we introduce the ``quantum rotation" $\tilde{r}$:
\begin{equation}
\tilde{r} = r+o
\end{equation}
For preons
\begin{equation}
\tilde{r}_p = o_p
\end{equation}
For the presently observed elementary fermions
\begin{equation}
\tilde{r} = r +1
\end{equation}
One may view the symmetry of an elementary particle, defined by representations of the SLq(2) algebra, in any of the following ways:
\begin{equation}
\mbox{D}_{mm'}^j = \mbox{D}_{-3t_3-3t_0}^{3t} = \mbox{D}_{\frac{w}{2}\frac{\tilde{r}}{2}}^{\frac{N}{2}} = \tilde{\mbox{D}}_{\nu_a \nu_b}^{N'}
\end{equation}
where $N'$ is the total number of preons. The quantum knot-preon complementary representations are related by
\begin{equation}
\tilde{\mbox{D}}_{\nu_a \nu_b}^{N'} = \sum_{N,w,r} \delta \left( N', N \right) \delta \left(\nu_a + \nu_b, w \right) \delta \left(\nu_a - \nu_b, \frac{\tilde{r}}{2} \right) \mbox{D}_{\frac{w}{2} \frac{\tilde{r}}{2}}^{\frac{N}{2}}
\end{equation}

The previous considerations are based on electroweak physics. To describe the strong interactions it is necessary according to the standard model to introduce SU(3). In the SLq(2) electroweak model the need for the additional SU(3) symmetry appears already at the level of leptons and neutrinos since these are presented in the SLq(2) model as $a^3$ and $c^3$ respectively. Then the simple way to satisfy the Pauli principle is to make the replacements:
\begin{displaymath}
	\begin{matrix}
	\text{leptons}: & a^3 \rightarrow  & \varepsilon^{ijk} a_i a_j a_k \\
	\text{neutrinos}: &c^3 \rightarrow & \varepsilon^{ijk}c_i c_j c_k
	\end{matrix}
\end{displaymath}
where $a_i$ and $c_i$ provide a basis for the fundamental representation of SU(3). Then the leptons and neutrinos are color singlets. If the $b$ and $d$ preons are also color singlets, then down quarks $a_ib^2$ and up quarks $c_id^2$ provide a basis for the fundamental representation of SU(3), as required by the standard model.

\newpage
\begin{center}
\underline{\textbf{Fig. 4.1}}
\end{center}
\begin{center}
\begin{tabular}{cc|cc}
 & \underline{$\left( w, r, o \right)$} & & \underline{$\left( w, r, o \right)$} \\
 Leptons, $\mbox{D}^{\frac{3}{2}}_{\frac{3}{2} \frac{3}{2}} \sim a^3$ & & $a$-preons, $\mbox{D}^{\frac{1}{2}}_{\frac{1}{2} \frac{1}{2}}$ \\
 
 \includegraphics[scale=0.45]{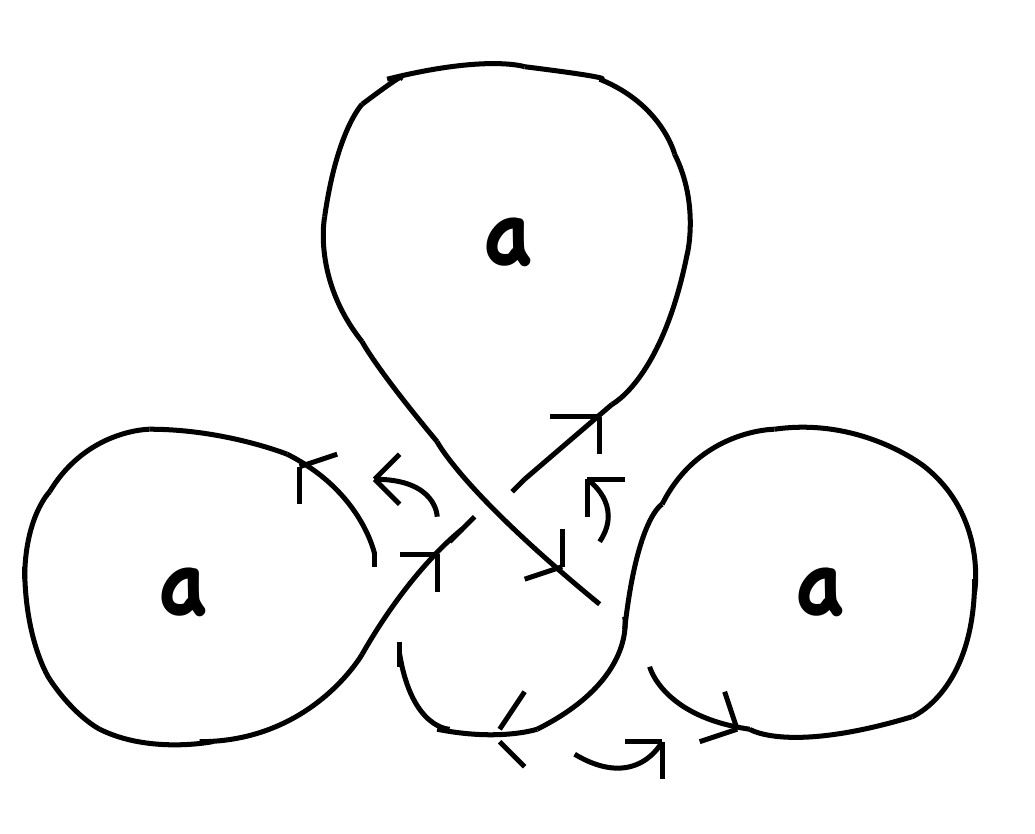} & $\left(3,2,1\right)$ & \includegraphics[scale=0.45]{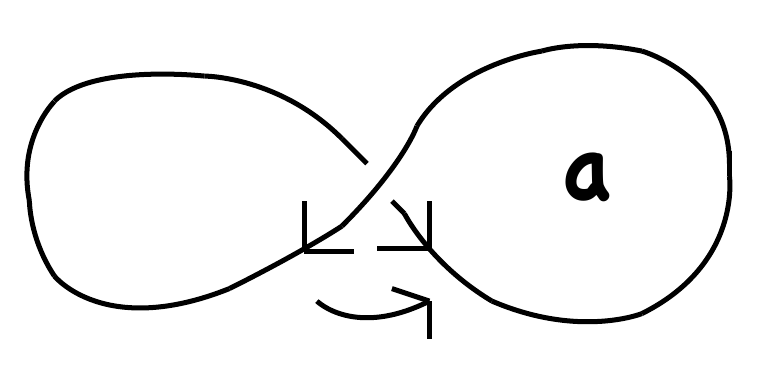} & $\left(1,0,1\right)$ \\
 \hline
 Neutrinos, $\mbox{D}^{\frac{3}{2}}_{-\frac{3}{2} \frac{3}{2}} \sim c^3$ & & $c$-preons, $\mbox{D}^{\frac{1}{2}}_{-\frac{1}{2} \frac{1}{2}}$ & \\
 
 \includegraphics[scale=0.45]{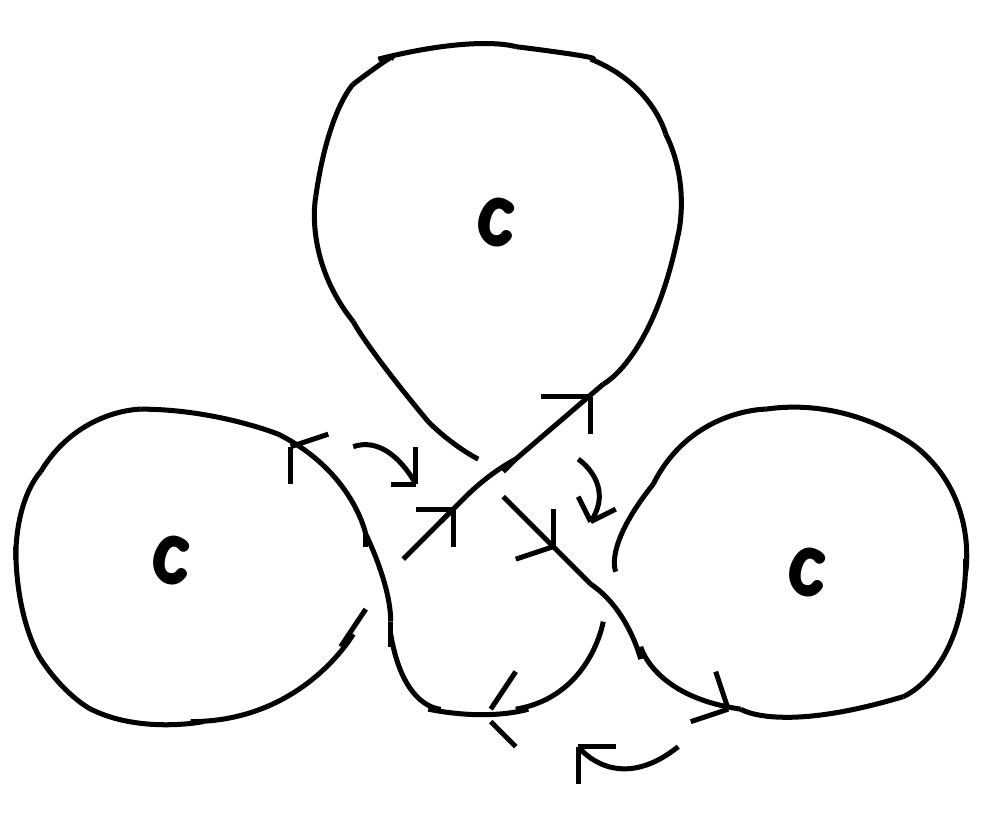} & $\left(-3,2,1 \right)$ & \includegraphics[scale=0.45]{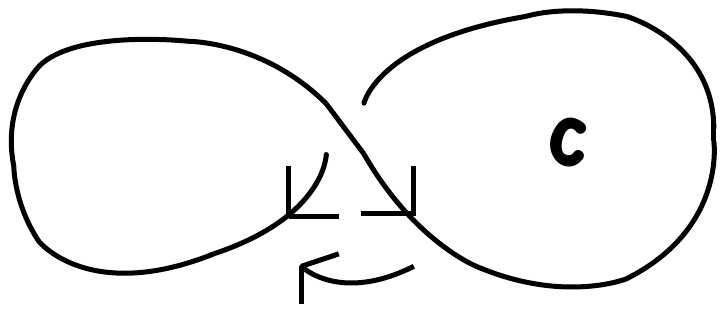} & $\left(-1,0,1 \right)$\\
 \hline
 $d$-quarks, $\mbox{D}^{\frac{3}{2}}_{\frac{3}{2} -\frac{1}{2}} \sim ab^2$ & & $b$-preons, $\mbox{D}^{\frac{1}{2}}_{\frac{1}{2}-\frac{1}{2}}$ & \\
 
\includegraphics[scale=0.45]{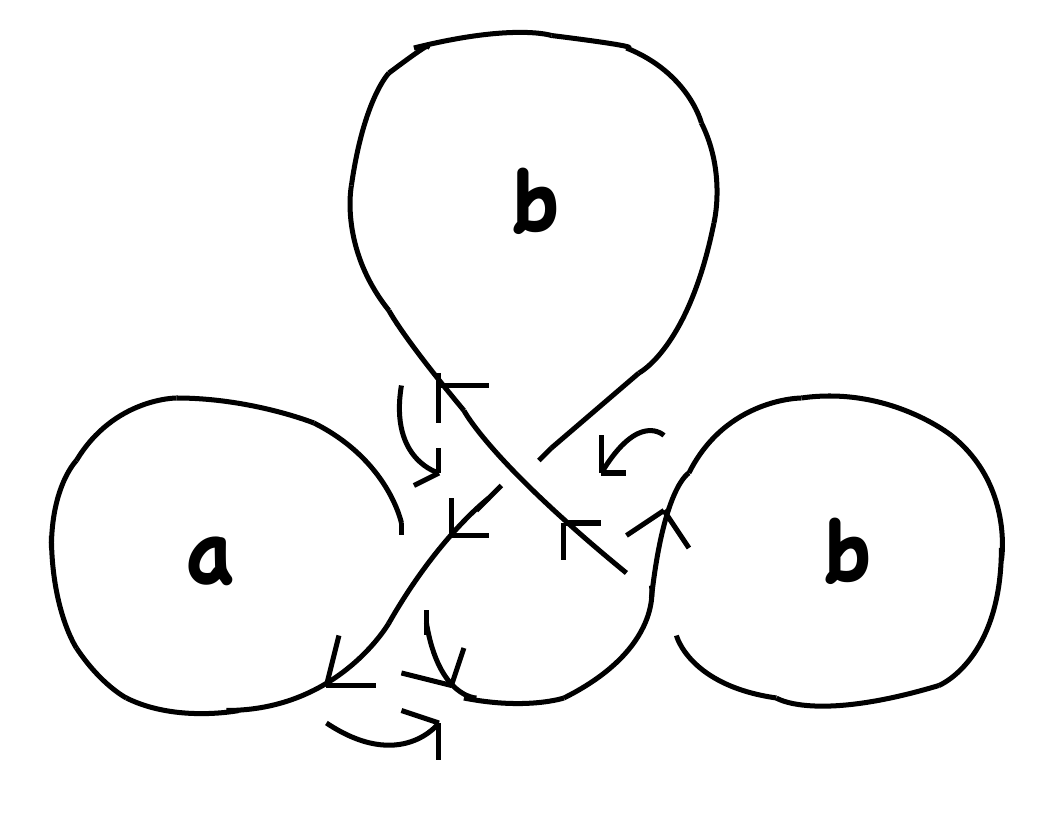} & $\left( 3,-2,1\right)$ & \includegraphics[scale=0.45]{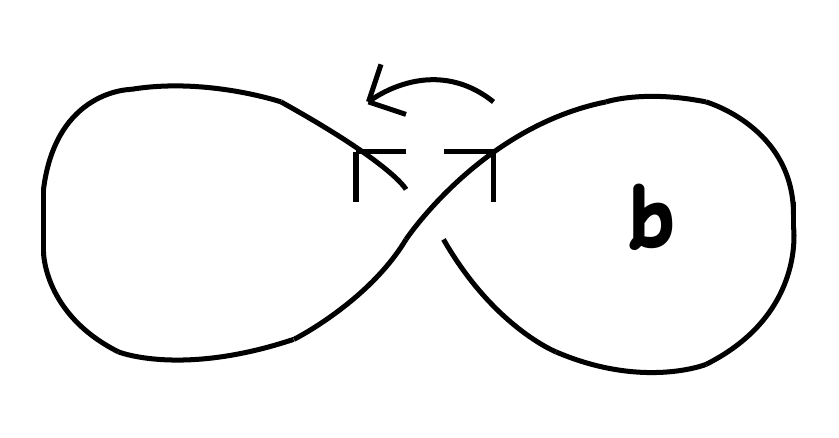} & $\left(1,0,-1\right)$ \\
\hline
$u$-quarks, $\mbox{D}^{\frac{3}{2}}_{-\frac{3}{2} -\frac{1}{2}} \sim cd^2$ & & $d$-preons, $\mbox{D}^{\frac{1}{2}}_{-\frac{1}{2} -\frac{1}{2}}$ \\

\includegraphics[scale=0.45]{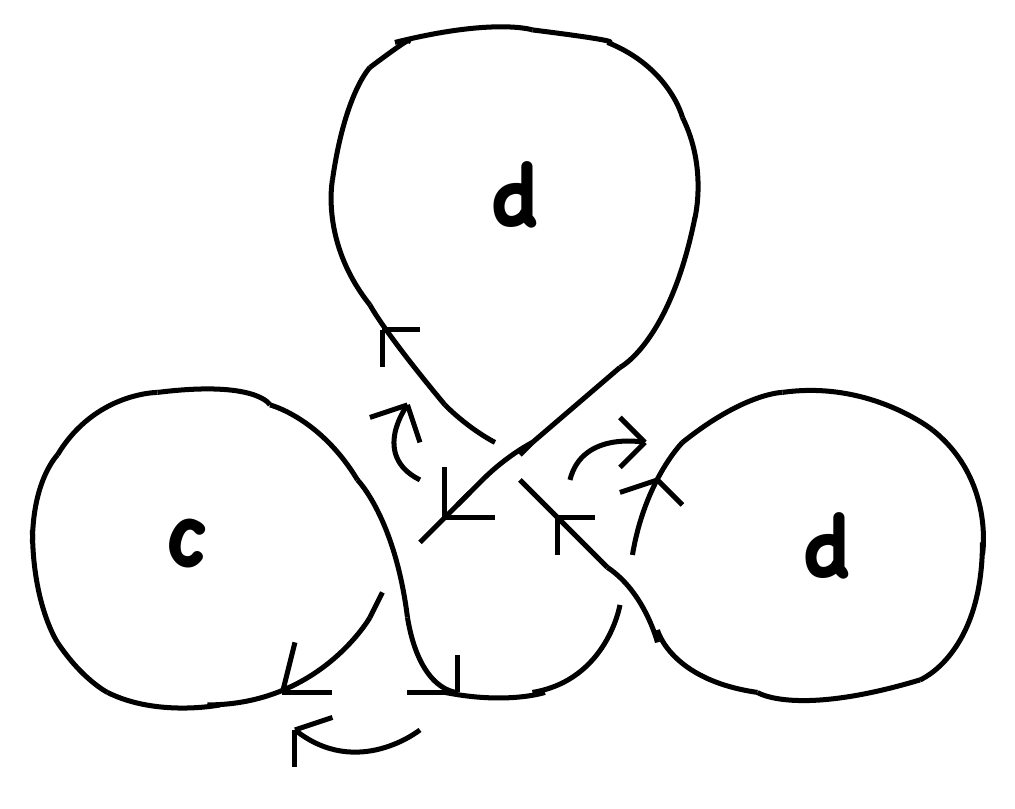} & $\left(-3,-2,1 \right)$ & \includegraphics[scale=0.45]{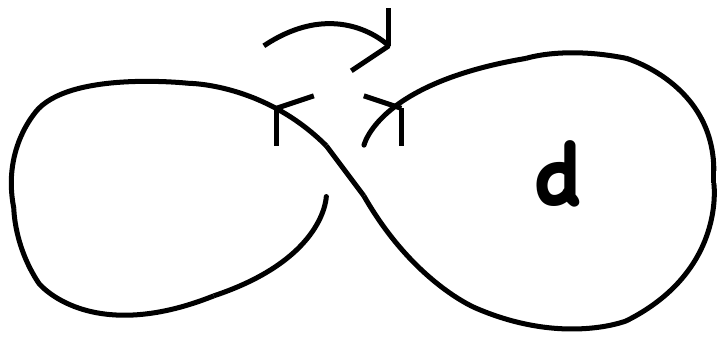} & $\left(-1,0,-1 \right)$ \\
\end{tabular}
\end{center}
\begin{center}
$Q = -\frac{e}{6} \left( w+r+o \right)$
\end{center}
\begin{center}
$(j,m,m') = \frac{1}{2}(N, w, r+o)$
\end{center} 

One may interpret the elements $(a,b,c,d)$ of the SLq(2) algebra as creation operators for either preonic particles or flux loops. Then the D$_{mp}^j$ may be interpreted as a creation operator for a composite particle composed of either preonic particles or flux loops. These two different views of the same particle may be reconciled as describing $N$-body systems bound by a knotted field having $N$-crossings with one preon at each crossing as illustrated in Figure 4.1 for $N=3$. In the limit where the three outside lobes become infinitesimal compared to the central circuit, the resultant structure will resemble a three particle system tied together by a Nambu-like string. Since the topological diagram of Fig. 4.1 describes loops that have no size or shape, one needs to introduce an explicit Lagrangian to go further.

\section{The Preon Lagrangian$^{(1)}$}

We shall assume that there is a common part of the knot Lagrangian shared by all representations but that there may be additional preonic terms. We shall describe only the common part of the knot Lagrangian by first recording the Lagrangian of the standard model in the following form:

\begin{equation}
\mathcal{L}_{st} = \mbox{\Rmnum{1}} + \mbox{\Rmnum{2}} + \mbox{\Rmnum{3}} + \mbox{\Rmnum{4}}
\end{equation}
where
\begin{displaymath}
\mbox{\Rmnum{1}} = \frac{1}{4}\mbox{Tr}\mbox{W}^{\mu \lambda} \mbox{W}_{\mu \lambda} - \frac{1}{4}\mbox{H}^{\mu \lambda} \mbox{H}_{\mu \lambda} \tag{5.1 \Rmnum{1}}
\end{displaymath}
\begin{displaymath}
\mbox{\Rmnum{2}} = i \left[ \bar{L} \nabla L + \bar{R} \nabla R    \right] \tag{5.1 \Rmnum{2}}
\end{displaymath}
\begin{displaymath}
\mbox{\Rmnum{3}} = \frac{1}{2} \left[ \bar{\nabla} \varphi \nabla \varphi - V(\bar{\varphi} \varphi)  \right] \tag{5.1 \Rmnum{3} }
\end{displaymath}
\begin{displaymath}
\mbox{\Rmnum{4}} = -\frac{m}{\rho} \left[ \bar{L} \varphi R + \bar{R} \varphi L \right] \tag{5.1 \Rmnum{4}}
\end{displaymath}
\setcounter{equation}{1}
To obtain the knot Lagrangian replace every field operator, $\Psi(x)$, of the standard model by a knotted field operator as follows:
\begin{equation}
\hat{\Psi}(x) = \Psi(x) \mbox{D}_{mm'}^j
\end{equation}
where $\Psi(x)$ transforms under SU(2)$\times$U(1) and D$_{mm'}^j$ transforms under U$_a(1) \times$U$_b(1)$. If $\hat{\Psi}(x)$ is a preonic field operator, we assume that the hypothetical $\Psi(x)$ is also a doublet transforming under SU(2)$\times$U(1) and that D$_{mm'}^j$ is a preonic representation of SLq(2).

We now replace the separate parts of $\mathcal{L}_{st}$, beginning with the Higgs mass term \Rmnum{4} describing the four preons. In the standard model $L$ and $\varphi$ are isotopic doublets while $\bar{L} \varphi$ and $R$ are isotopic singlets. We retain this isotopic structure and continue to follow the standard model by going to the unitary gauge where $\varphi$ has a single component which is neutral. In passing to the SLq(2) algebra we assume that $\varphi$ is a SLq(2) singlet and that $L$ and $R$ carry the same D$_{mm'}^j$.

The $(\nu, \ell)$ and $(u, d)$ fermions are $t_3$-doublets in Table 2.1. They transform under the $t=1/2$ representation of the isotopic spin and lie in the $j=3/2$ representation of SLq(2). By Table (3.1) the $(c, a)$ and $(d, b)$ preons, lying in the $j=1/2$ representation of SLq(2), are also $t_3$-doublets but transform under U$_a \times$ U$_b$. Then by (5.2) and Table 2.1 the composite fermion $t_3$-doublets are
\begin{equation}
L(\nu, \ell) = \begin{pmatrix} \Psi_L(\nu) \mbox{D}_{-\frac{3}{2}\frac{3}{2}}^{^{3/2}} \\ \Psi_L (\ell) \mbox{D}_{\frac{3}{2}\frac{3}{2}}^{^{3/2}} \end{pmatrix} \rightarrow \begin{pmatrix} \Psi_L(\nu) c^3 \\ \Psi_L(\ell) a^3 \end{pmatrix}
\end{equation}
\begin{equation}
L(u, d) = \begin{pmatrix} \Psi_L(u) \mbox{D}_{-\frac{3}{2}-\frac{1}{2}}^{^{3/2}} \\ \Psi_L (d) \mbox{D}_{\frac{3}{2}-\frac{1}{2}}^{^{3/2}} \end{pmatrix} \rightarrow \begin{pmatrix} \Psi_L(u) cd^2 \\ \Psi_L(d) ab^2 \end{pmatrix}
\end{equation}
where $(\Psi_L(\nu), \Psi_L(\ell) )$ and $(\Psi_L(u), \Psi_L(d))$ are $t_3$-doublets in the $t=1/2$ representation of SU(2) while $(c^3, a^3)$ and $(cd^2, ab^2)$ are, by Table 3.3, $t_3$-doublets in the $j=3/2$ representation of SLq(2).

The corresponding preon doublets are, by Tables 3.1 and 3.2
\begin{equation}
L(c, a) = \begin{pmatrix} \Psi_L(c) \mbox{D}_{-\frac{1}{2}\frac{1}{2}}^{^{1/2}} \\ \Psi_L (a) \mbox{D}_{\frac{1}{2}\frac{1}{2}}^{^{1/2}} \end{pmatrix} \rightarrow \begin{pmatrix} \Psi_L(c) c \\ \Psi_L(a) a \end{pmatrix}
\end{equation}
\begin{equation}
L(d, b) = \begin{pmatrix} \Psi_L (d) \mbox{D}_{-\frac{1}{2}-\frac{1}{2}}^{^{1/2}} \\ \Psi_L(b) \mbox{D}_{\frac{1}{2}-\frac{1}{2}}^{^{1/2}}  \end{pmatrix} \rightarrow \begin{pmatrix}\Psi_L(d) d \\ \Psi_L(b) b  \end{pmatrix}
\end{equation}
where the hypothetical $(\Psi_L(c), \Psi_L(a))$ and $(\Psi_L(d), \Psi_L(b))$ are also assumed to be $t_3$-doublets in the $t=1/2$ representation of SU(2) while $(c,a)$ and $(d,b)$ are by Table 3.1 $t_3$-doublets in the $j=1/2$ representation of SLq(2).

In constructing both the knot and the preon Lagrangians we make the basic assumption that $\Psi(x)$ in $(5.2-5.6)$ is a local SU(2) field like the field operators of the standard model. In both the knot and preon actions therefore, the Lagrangians express the dynamics of the standard model and differ from the standard model and each other only by form factors dependent on $\bar{\mbox{D}}_{mm'}^j \mbox{D}_{mm'}^j$. These form factors are U$_a\times$U$_b$ invariant and are functions of SLq(2) parameters. The differences between the preon and knot Lagrangians then stem from the form factors of the $j=1$ vectors and $j=1/2$ fermions in the preon Lagrangian since these differ from the form factors of the $j=3$ vectors and $j=3/2$ fermions in the knot Lagrangian. In both the knot and preon actions, charge and hypercharge are recorded by D$_{mm'}^j$ and their conservation follows from the invariance of the action under U$_a(1) \times$U$_b(1)$ transformations.

Let us now begin to interpret $(5.1)$ as a preon Lagrangian by first expressing the Higgs mass term as a sum of four parts, the contributions of the four preons, as follows:
\begin{equation}
\mbox{\Rmnum{4}} = \mbox{\Rmnum{4}}_a + \mbox{\Rmnum{4}}_b + \mbox{\Rmnum{4}}_c + \mbox{\Rmnum{4}}_d + \mbox{adjoint}
\end{equation}
where

\begin{equation}
\begin{array}{lcr}
\mbox{\Rmnum{4}}_a = \bar{L}(c,a) \Phi_a R(a) & \hspace{0.3cm}& \mbox{\Rmnum{4}}_b = \bar{L}(d,b) \Phi_b R(b)
\end{array}
\end{equation}
\begin{equation}
\begin{array}{lcr}
\mbox{\Rmnum{4}}_c = \bar{L}(c,a)\Phi_c R(c) & \hspace{0.3cm} & \mbox{\Rmnum{4}}_d = \bar{L}(d,b)\Phi_d R(d)
\end{array}
\end{equation}
where $L(c,a)$ and $L(d,b)$ are given by (5.5) and (5.6), respectively.
Here the Higgs doublets are
\begin{equation}
\begin{array}{lcr}
\Phi_a = \begin{pmatrix} 0 \\ \rho_a \end{pmatrix} & \hspace{0.8cm} & \Phi_b = \begin{pmatrix} 0 \\ \rho_b \end{pmatrix}
\end{array}
\end{equation}
\begin{equation}
\begin{array}{lcr}
\Phi_c = \begin{pmatrix} \rho_c \\ 0 \end{pmatrix} & \hspace{0.8cm} & \Phi_d = \begin{pmatrix} \rho_d \\ 0 \end{pmatrix}
\end{array}
\end{equation}
and $L$ and $R$ carry the same knot factors, D$_{mp}^{^{1/2}}$.
Then one may compute the $a$-contribution to $(5.7)$ as follows:
\begin{eqnarray}
\mbox{\Rmnum{4}}_a &=& \begin{pmatrix} \bar{\Psi}_L(c) \bar{c}\hspace{0.1cm}, & \bar{\Psi}_L (a) \bar{a} \end{pmatrix}
\begin{pmatrix} 0 \\ \rho_a \end{pmatrix} \begin{pmatrix} \Psi_R (a) a \end{pmatrix} \nonumber \\
&=& \rho_a \bar{\Psi}_L(a) \Psi_R(a) \bar{a}a 
\end{eqnarray}
The adjoint is
\begin{eqnarray}
\overline{\mbox{\Rmnum{4}}}_a &=& \begin{pmatrix} \bar{\Psi}_R (a) \bar{a} \end{pmatrix}\begin{pmatrix} 0\hspace{0.1cm}&\rho_a \end{pmatrix} \begin{pmatrix} \Psi_L(c) c \\ \Psi_L(a)a \end{pmatrix} \nonumber \\
&=& \rho_a \bar{\Psi}_R(a) \Psi_L(a) \bar{a}a
\end{eqnarray}
Then
\begin{equation}
\mbox{\Rmnum{4}}_a + \overline{\mbox{\Rmnum{4}}}_a = \rho_a \bar{a}a \bar{\Psi}(a) \Psi(a)
\end{equation}

Replacements like $(5.3)-(5.6)$ that define the knot and preon Lagrangian lead to expressions like $(5.14)$ and therefore to an operator replacement of the Lagrangian of the standard model in terms of the preonic operators $(a,b,c,d)$. The physical knot Lagrangian itself is interpreted in terms of the physical state space on which $(a,b,c,d)$ can operate. We may take the eigenstates of $b$ and $c$ to be a basis in this space and for the physical interpretation we may take these states to be either states of particles or states of flux loops. We may restrict the dimensionality of the physical state space to three generations of the $t=1/2$ fermions and to a single generation denoted by $\ket{0}$ of the hypothetical $t=1/3$ preons.

The $a$-contribution is then given by $(5.14)$
\begin{equation}
\bra{0} \mbox{\Rmnum{4}}_a + \overline{\mbox{\Rmnum{4}}}_a \ket{0} = m_a \bar{\Psi}(a) \Psi(a)
\end{equation}
where the mass of the $a$-preon is
\begin{equation}
m_a = \rho_a \bra{0} \bar{a}a \ket{0}
\end{equation}
The masses of all the preons are obtained in the same way with the result
\begin{equation}
m_p = \rho_p \bra{0} \bar{p}p \ket{0} \hspace{1.0cm} p =a,b,c,d
\end{equation}
where $\ket{0}$ is the ground state of the $b$ and $c$ elements
\begin{eqnarray}
b \ket{0} &=& \beta \ket{0} \\ c \ket{0} &=& \gamma \ket{0}
\end{eqnarray}
In (5.17) $\bra{0} \bar{p}p \ket{0}$ may be reduced to a simple function of $q, \beta$, and $\gamma$ by going to the unitary algebra, SUq(2), where $\overline{a} = d$ and $\overline{b} = -qc$.$^{(1)}$

\newpage
\section{The Fermion-Boson Interaction}

The interaction $(5.1 \hspace{0.1cm} \mbox{\Rmnum{2}})$ is
\begin{equation}
\mbox{\Rmnum{2}} = i \left[ \bar{L} \nabla L + \bar{R} \nabla R \right]
\end{equation}
where $L$ and $R$ are the left and right chiral fields, $L$ is the isotopic doublet described in $(5.3)$ and $(5.4)$ at the composite fermionic level, or by $(5.5)$ and $(5.6)$ at the preonic level. $R$ is an isotopic singlet with the same knot factor as $L$. Here $\nabla$ is the covariant derivative
\begin{equation}
\nabla = \slashed{\partial} + \slashed{\mathcal{W}}
\end{equation}
where $\mathcal{W}$ is the vector connection.

\noindent If $\mathcal{W}$ is the vector connection of the standard model, then
\begin{equation}
\mathcal{W} = i g \left( \mbox{W}^+t_+ + \mbox{W}^-t_- + \mbox{W}^3 t_3 \right) + ig_0 \mbox{W}^0 t_0
\end{equation}
To go over to the knot and preon models we follow $(2.13)$ and therefore replace $(\mbox{W}^+, \mbox{W}^-, \mbox{W}^3)$ in $(6.3)$ by $(\mbox{W}^+\mbox{D}_{-30}^3, \mbox{W}^-\mbox{D}_{30}^3, \mbox{W}^3\mbox{D}_{00}^3)$ in the $j=3$ representation of SLq(2) and by \newline $\left( \mbox{w}^+\mbox{D}_{-10}^1, \mbox{w}^-\mbox{D}_{10}^1, \mbox{w}^0\mbox{D}_{00}^1\right)$ in the $j=1$ representation of SLq(2). Alternatively we may replace $\vec{t}$ in $(6.3)$ by $\vec{\tau}$ as follows:
\begin{eqnarray}
\tau_{\pm}^j &=& c_{\pm}^jt_{\pm}\mathcal{D}_{\pm}^j \hspace{3.0cm}  j = 1, 3 \\
\tau_3^j &=& c_3^j t_3 \mathcal{D}_3^j \hspace{3.22cm} j = 1,3
\end{eqnarray}
Here $\mathcal{D}_{\pm}^j$ is the operator part of the monomial D$_{\pm}^j$, shown as follows:
\begin{eqnarray}
\mathcal{D}_{\pm}^j &=& \mbox{D}_{\pm}^j / \mbox{A}_{\pm j 0}^j \hspace{2.1cm} j = 1, 3
\end{eqnarray}
and
\begin{eqnarray}
\mathcal{D}_3^j &=& \mbox{D}_{00}^j \hspace{3.09cm} j=1, 3
\end{eqnarray}

\noindent D$_{00}^j$ is not a monomial but is an operator function of $bc$.

\noindent The determination of the constants $c_k^j$ in (6.4) and (6.5) is deferred to the next section where they are fixed by the Higgs kinetic energy.

\noindent We also assume in both the $j=3$ and $j=1$ representations
\begin{eqnarray}
(j,m,p) &=& 3 (t, -t_3, -t_0) \\
Q &=& e(t_3+t_0)
\end{eqnarray}

In the adjoint representation where $j=1$ and $t=+1/3$, the possible values of $m$ and $p$ are $(1,0,-1)$ and the corresponding values for $t_3$ and $t_0$ are $(1/3, 0, -1/3)$ by $(6.8)$. We assign $t_0 = 0$ to all the vector bosons. Then we have by $(6.9)$ the Tables (6.1) and (6.2) for the $j=3$ (knotted electroweak) and $j=1$ (adjoint vectors).

\begin{center}
{\underline{{\bf{Adjoint Vector}}}}
\end{center}
\begin{center}
\begin{tabular}{c | c c c c l}
  & $t$ & $t_3$ &  $t_0$ & $Q$ &  $\mbox{D}^{3t}_{-3t_3-3t_0}$ \\
 \hline
 w$^+$ & $\frac{1}{3}$ & $\frac{1}{3}$ & 0 & $\frac{e}{3}$ &  $\mbox{D}^{1}_{-10}=cd = \mathcal{D}_+\left(\frac{1}{3}\right)$ \\
 w$^-$ & $\frac{1}{3}$ & -$\frac{1}{3}$ & 0 &  -$\frac{e}{3}$ &  $\mbox{D}^{1}_{10}=ab = \mathcal{D}_-\left(\frac{1}{3}\right)$\\
 w$^3$ & $\frac{1}{3}$ & 0 & 0 & 0 &  $\mbox{D}^{1}_{00}=ad+bc = \mathcal{D}_0 \left( \frac{1}{3} \right)$\\
\end{tabular}
\end{center}
\begin{center}
\bf{Table 6.1}
\end{center}

\begin{center}
{\underline{{\bf{Knotted Electroweak Vector}}}}
\end{center}
\begin{center}
\begin{tabular}{c | c c c c l}
  & $t$ & $t_3$ &  $t_0$ & $Q$ &  $\mbox{D}^{3t}_{-3t_3-3t_0}$ \\
 \hline
 W$^+$ & $1$ & $1$ & 0 & $e$ &  $\mbox{D}^{3}_{-30}\sim c^3d^3 = \mathcal{D}_+\left(1\right)$ \\
 W$^-$ & $1$ & -$1$ & 0 &  -$e$ &  $\mbox{D}^{3}_{30} \sim a^3b^3 = \mathcal{D}_-\left(1\right)$\\
 W$^3$ & $1$ & 0 & 0 & 0 &  $\mbox{D}^{3}_{00}=f_3\left(bc\right) = \mathcal{D}_0 \left( 1 \right)$\\
\end{tabular}
\end{center}
\begin{center}
\bf{Table 6.2}
\end{center}
The left chiral interaction terms are by $(5.1 \hspace{0.1cm} \mbox{\Rmnum{2}})$
\begin{equation}
\overline{L} \nabla L = \mbox{\Rmnum{2}} (c,a) + \mbox{\Rmnum{2}} (d,b)
\end{equation}
where
\begin{equation}
\mbox{\Rmnum{2}}(c,a) = \bra{0} \begin{pmatrix} \overline{\Psi}_L(c) \bar{c}\hspace{0.1cm}, & \overline{\Psi}_L(a) \bar{a} \end{pmatrix}\begin{pmatrix} \slashed{\partial} + \slashed{\mathcal{W}} \end{pmatrix}\begin{pmatrix} \Psi_L(c)c \\ \Psi_L(a)a \end{pmatrix} \ket{0}
\end{equation}
and
\begin{equation}
\mbox{\Rmnum{2}}(d,b) = \bra{0} \begin{pmatrix} \overline{\Psi}_L(d) \bar{d} \hspace{0.1cm}, & \overline{\Psi}_L(b)\bar{b} \end{pmatrix} \begin{pmatrix} \slashed{\partial} + \slashed{\mathcal{W}} \end{pmatrix} \begin{pmatrix} \Psi_L(d) d \\ \Psi_L(b)b \end{pmatrix} \ket{0}
\end{equation}
In (6.11) one has
\begin{displaymath}
\bra{0} \begin{pmatrix} \overline{\Psi}_L(c) \bar{c}\hspace{0.1cm}, & \overline{\Psi}_L(a) \bar{a}  \end{pmatrix} \slashed{\partial} \begin{pmatrix}  \Psi_L(c) c \\ \Psi_L(a)a \end{pmatrix} \ket{0} \hspace{7.0cm}
\end{displaymath}
\begin{eqnarray}
&=& \bra{0} \bar{c}c\ket{0} \overline{\Psi}_L(c) \slashed{\partial} \Psi_L(c) + \bra{0} \bar{a}a \ket{0} \overline{\Psi}_L(a) \slashed{\partial} \Psi_L(a) \\
&=& \overline{\Psi}_L(c) \slashed{\Delta}_c \Psi_L(c) + \overline{\Psi}_L(a) \slashed{\Delta}_a \Psi_L(a)
\end{eqnarray}
where $\slashed{\Delta}_c$ and $\slashed{\Delta}_a$ are the following rescaled momentum operators
\begin{displaymath}
\slashed{\Delta}_c = \bra{0} \bar{c}c \ket{0} \slashed{\partial} \tag{6.15$c$}
\end{displaymath}
and
\begin{displaymath}
\slashed{\Delta}_a = \bra{0} \bar{a}a \ket{0} \slashed{\partial} \tag{6.15$a$}
\end{displaymath}
\setcounter{equation}{15}

\noindent The momentum operators are rescaled in the same way as the Higgs rest masses:
\begin{displaymath}
	m_p = \bra{0} \bar{p} p \ket{0} \hskip1.5cm p=(a,b,c,d)
\end{displaymath}
From (6.11) and $(6.3)-(6.5)$ one also has
\begin{displaymath}
\bra{0} \begin{pmatrix} \overline{\Psi}_L(c) \bar{c} \hspace{0.1cm}, & \overline{\Psi}_L(a) \bar{a} \end{pmatrix} \mathcal{W} \begin{pmatrix} \Psi_L(c)c \\ \Psi_L(a)a \end{pmatrix} \ket{0} \hspace{7.0cm}
\end{displaymath}
\begin{eqnarray}
&=& \bra{0} \begin{pmatrix} \overline{\Psi}_L(c)\bar{c} \hspace{0.1cm}, & \overline{\Psi}_L(a) \bar{a} \end{pmatrix} \begin{pmatrix} c_3 \mathcal{D}_3 \mbox{w}^3 & c_+ \mathcal{D}_+ \mbox{w}^+ \\ c_- \mathcal{D}_- \mbox{w}^- & -c_3 \mathcal{D}_3 \mbox{w}^3 \end{pmatrix} \begin{pmatrix} \Psi_L(c) c \\ \Psi_L(a)a \end{pmatrix} \ket{0} \\
&=& \mbox{F}_{\bar{c}c} \left[ \hspace{0.05cm}\overline{\Psi}_L (c) \mbox{w}^3 \Psi_L(c) \right] + \mbox{F}_{\bar{c}a} \left[ \hspace{0.05cm} \overline{\Psi}_L(c) \mbox{w}^+ \Psi_L(a) \right] \nonumber \\
&& \hspace{0.3cm}+ \mbox{F}_{\bar{a}c} \left[ \hspace{0.05cm} \overline{\Psi}_L (a) \mbox{w}^- \Psi_L(c) \right] - \mbox{F}_{\bar{a}a} \left[ \hspace{0.05cm} \overline{\Psi}_L (a) \mbox{w}^3 \Psi_L(a) \right]
\end{eqnarray}
Here w$^+$, w$^-$, and w$^3$ are components of the adjoint vector field ($j=1$) that mediate the interaction between the left chiral preons. The adjoint vector-preon form factors are
\begin{equation}
\begin{array}{rcl}
\mbox{F}_{\bar{c}c} & = & c_3 \bra{0} \bar{c}\hspace{0.05cm} \mathcal{D}_3 c \ket{0} \\ & = & c_3 \bra{0} \bar{c} f_3(bc) c \ket{0}
\end{array}
\hspace{1.0cm}
\begin{array}{rcl}
\mbox{F}_{\bar{c}a} & = & c_+ \bra{0} \bar{c} \hspace{0.05cm} \mathcal{D}_+ a \ket{0} \\ & = & c_+ \bra{0} \bar{c} \left(cd\right) a \ket{0} 
\end{array}
\end{equation}
\begin{equation}
\begin{array}{rcl}
\mbox{F}_{\bar{a}c} & = & c_- \bra{0} \bar{a}\hspace{0.05cm} \mathcal{D}_- c \ket{0} \\ & = & c_- \bra{0} \bar{a} \left(ab \right) c \ket{0}
\end{array}
\hspace{1.0cm}
\begin{array}{rcl}
\mbox{F}_{\bar{a}a} & = & c_3 \bra{0} \bar{a} \hspace{0.05cm} \mathcal{D}_3 a \ket{0} \\ & = & c_3 \bra{0} \bar{a} f_3(bc) a \ket{0} 
\end{array}
\end{equation}
by (6.7) and Table 6.1 where
\begin{equation}
f_3(bc) = ad+bc
\end{equation}
The form factors (6.18) and (6.19) are all invariant under U$_a \times $U$_b$ and are therefore expressible as functions of the knot parameters.

\noindent From (6.12) one has
\begin{displaymath}
\bra{0} \begin{pmatrix} \overline{\Psi}_L(d) \bar{d}\hspace{0.1cm}, & \overline{\Psi}_L(b) \bar{b}  \end{pmatrix} \slashed{\partial} \begin{pmatrix}  \Psi_L(d) d \\ \Psi_L(b)b \end{pmatrix} \ket{0} \hspace{7.0cm}
\end{displaymath}
\begin{eqnarray}
&=& \bra{0} \bar{d}d\ket{0} \overline{\Psi}_L(d) \slashed{\partial} \Psi_L(d) + \bra{0} \bar{b}b \ket{0} \overline{\Psi}_L(b) \slashed{\partial} \Psi_L(b) \nonumber  \\
&=& \overline{\Psi}_L(d) \slashed{\Delta}_d \Psi_L(d) + \overline{\Psi}_L(b) \slashed{\Delta}_b \Psi_L(b)
\end{eqnarray}
where $\slashed{\Delta}_d$ and $\slashed{\Delta}_b$ are rescaled momentum operators
\begin{displaymath}
\slashed{\Delta}_d = \bra{0} \bar{d}d \ket{0} \slashed{\partial} \tag{6.22$d$}
\end{displaymath}
and
\begin{displaymath}
\slashed{\Delta}_b = \bra{0} \bar{b}b \ket{0} \slashed{\partial} \tag{6.22$b$}
\end{displaymath}
\setcounter{equation}{22}
From (6.12) one also has (again excluding $\mbox{w}^0t_0$)
\begin{displaymath}
\bra{0} \begin{pmatrix} \overline{\Psi}_L(d) \bar{d} \hspace{0.1cm}, & \overline{\Psi}_L(b) \bar{b} \end{pmatrix} \mathcal{W} \begin{pmatrix} \Psi_L(d)d \\ \Psi_L(b)b \end{pmatrix} \ket{0} \hspace{7.0cm}
\end{displaymath}
\begin{eqnarray}
&=& \bra{0} \begin{pmatrix} \overline{\Psi}_L(d)\bar{d} \hspace{0.1cm}, & \overline{\Psi}_L(b) \bar{b} \end{pmatrix} \begin{pmatrix} c_3 \mathcal{D}_3 \mbox{w}^3 & c_+ \mathcal{D}_+ \mbox{w}^+ \\ c_- \mathcal{D}_- \mbox{w}^- & -c_3 \mathcal{D}_3 \mbox{w}^3 \end{pmatrix} \begin{pmatrix} \Psi_L(d) d \\ \Psi_L(b)b \end{pmatrix} \ket{0} \\
&=& \mbox{F}_{\bar{d}d} \left[ \hspace{0.05cm}\overline{\Psi}_L (d) \mbox{w}^3 \Psi_L(d) \right] + \mbox{F}_{\bar{d}b} \left[ \hspace{0.05cm} \overline{\Psi}_L(d) \mbox{w}^+ \Psi_L(b) \right] \nonumber \\
&& \hspace{0.3cm}+ \mbox{F}_{\bar{b}d} \left[ \hspace{0.05cm} \overline{\Psi}_L (b) \mbox{w}^- \Psi_L(d) \right] - \mbox{F}_{\bar{b}b} \left[ \hspace{0.05cm} \overline{\Psi}_L (b) \mbox{w}^3 \Psi_L(b) \right]
\end{eqnarray}
where 
\begin{equation}
\begin{array}{rcl}
\mbox{F}_{\bar{d}d} & = & c_3\bra{0} \bar{d} \hspace{0.05cm} \mathcal{D}_3 d \ket{0} \\ \mbox{F}_{\bar{b}d} & = & c_-\bra{0} \bar{b} \hspace{0.05cm} \mathcal{D}_- d \ket{0} 
\end{array}
\hspace{1.0cm}
\begin{array}{rcl}
\mbox{F}_{\bar{d}b} & = & c_+\bra{0}\bar{d} \hspace{0.05cm} \mathcal{D}_+ b \ket{0} \\ \mbox{F}_{\bar{b}b} & = & c_3\bra{0} \bar{b} \hspace{0.05cm} \mathcal{D}_3 b \ket{0} 
\end{array}
\end{equation}
The form factors
\begin{equation}
\bar{d} \hspace{0.05cm} \mathcal{D}_+ b = \bar{d} \left( c d \right) b
\end{equation}
and
\begin{equation}
\bar{b} \hspace{0.05cm} \mathcal{D}_- d = \bar{b} \left(ab \right) d
\end{equation}
are invariant under U$_a \times$ U$_b$ as required.

\noindent All the invariant form factors are simple functions of $\beta \gamma$.

\section{The Higgs Kinetic Energy Terms}

In both the standard model and the knot model we assume that the Higgs is coupled to the observed electroweak field. In the preon model we shall assume that the Higgs is also coupled to the hypothetical electroweak field described by the adjoint representation of SLq(2). Then there are two components of the covariant derivative of the Higgs as follows:

\begin{equation}
\nabla^j \hspace{0.05cm} \varphi \ket{0} = \partial \hspace{0.05cm} \varphi \ket{0} + i g \left[ \left( \mbox{W}^+ \tau_+ \right)^j + \left( \mbox{W}^- \tau_- \right)^j +\frac{\left(\mbox{Z}\hspace{0.05cm} \tau_0 \right)^j}{\cos \theta} \right] \varphi \ket{0} \hspace{1.0cm} j=1,3
\end{equation}
where $j$ refers to the representation and $\theta$ is the Weinberg angle. The corresponding contributions to the total kinetic energy are
\begin{equation}
\frac{1}{2}\bra{0}\mbox{Tr}\hspace{0.1cm} \overline{\nabla_{\mu}\varphi} \nabla^{\mu} \varphi \ket{0}^j = \frac{1}{2} \partial_{\mu} \hspace{0.05cm} \rho \hspace{0.05cm} \partial^{\mu} \rho + g^2 \rho^2 \left[ \mbox{I}_{++}^j\left( \mbox{W}^{\mu}_+ \mbox{W}_{+ \mu}\right)^{j} + \mbox{I}_{--}^j \left(\mbox{W}^{\mu}_-\mbox{W}_{-\mu}\right)^j + \frac{\mbox{I}_{33}^j}{\cos^2\theta} \left( \mbox{Z}^{\mu}\mbox{Z}_{\mu}\right)^j \right]
\end{equation}
Here
\begin{displaymath}
\varphi = \left( \begin{array}{c} 0 \\ \rho \\ \end{array} \right) 
\end{displaymath}
and
\begin{displaymath}
I^j_{kk} = \frac{1}{2} \mbox{Tr}  \bra{0} \overline{\tau}_k^j \hspace{0.1cm} \tau_k^j \ket{0}  \hspace{2.0cm} \begin{array}{rcl}
k & = & +,-,3 \\  j & = & 1,3
\end{array}
\end{displaymath}
and $\tau_k^j$ are given by (6.4) and (6.5).

To agree with the observed masses of the $\mbox{W}_+$, $\mbox{W}_-$, and $\mbox{Z}$ of the standard theory we have set $\bra{0} 0 \rangle =1$ and 
\begin{equation}
\mbox{I}^3_{kk} =1
\end{equation}
As the simplest generalization of (7.3) we shall now set
\begin{equation}
I^j_{kk} =1 \hspace{2.0cm} j =1,3
\end{equation}
Then
\begin{equation}
\left| c_k^j \right|^{-2} = \frac{1}{2} \bra{0} \overline{\mathcal{D}}_k^j \hspace{0.05cm} \mathcal{D}_k^j \ket{0} \hspace{2.0cm} \begin{array}{rcl}
j & = & 1,3 \\
k & = & +,-,3
\end{array}
\end{equation}
The expressions $\overline{\mathcal{D}}_k^j \hspace{0.05cm} \mathcal{D}_k^j$ are invariant under U$(a)\times$U$(b)$ and are therefore functions of $bc$. The explicit expressions for $\left( \mathcal{D}_+^j, \mathcal{D}_-^j, \mathcal{D}_3^j \right)$ are $\left(c^3d^3,a^3b^3, \mathcal{D}_{00}^3 \right)$ for $j=3$, and $\left( cd, ab, \mathcal{D}_{00}^1 \right)$ for $j=1$. The $| c_k^j|$ are then determined as functions of $\beta \gamma$.


\section{The Field Invariant}

We replace the field invariant of the standard model by 
\begin{equation}
\sum_j \bra{0} \mbox{Tr} \hspace{0.05cm}\mathcal{W}_{\mu \lambda}^j \hspace{0.05cm} \mathcal{W}^{j \mu \lambda} \ket{0}
\end{equation}
where $\mathcal{W}_{\mu \lambda}^j$ are the field strengths of the knot and preon models and where $\ket{0}$ is the ground state of the commuting $b$ and $c$ operators. The $j$-sum is over the two vector representations: $j=1,3$.
In (8.1) we have
\begin{equation}
\mathcal{W}_{\mu \lambda}^j = \left[ \nabla_{\mu}^j \hspace{0.1cm}, \hspace{0.05cm} \nabla_{\lambda}^j \right] \hspace{2.0cm} j=1,3
\end{equation}
where
\begin{equation}
\nabla_{\mu}^j = \partial_{\mu} + \mathcal{W}_{\mu}^j
\end{equation}
and
\begin{equation}
\mathcal{W}_{\mu}^{j} = ig \left( \mbox{W}_{\mu}^+ \tau_+^j + \mbox{W}^-_{\mu} \tau_-^j + \mbox{W}_{\mu}^3 \tau_3^j \right) \hspace{2.0cm} j=1,3
\end{equation}
Then by (8.2) and (8.3)
\begin{equation}
\mathcal{W}_{\mu \lambda}^j = i g \left( \partial_{\mu} \mbox{W}_{\lambda}^k - \partial_{\lambda}\mbox{W}_{\mu}^k \right) \tau_k^j - g^2\hspace{0.05cm} \mbox{W}_{\mu}^k \hspace{0.05cm} \mbox{W}_{\lambda}^{\ell} \left[ \tau_k^j \hspace{0.1cm}, \tau_{\ell}^j \right] \hspace{1.5cm} j=1,3 
\end{equation}
where $\left( k ,\ell \right) = \left( +,-,0 \right)$ and differs from the standard model by the substitution of $\tau_k^j$ for $t_k$.

The $\tau$-commutators introduce structure coefficients invariant under the gauge transformations U$_a(1) \times$U$_b(1)$ that leave the SLq(2) algebra invariant and hence are functions of $bc$ only. The structure coefficients in $(8.5)$ evaluated on $\ket{0}$ will therefore be functions of $\beta \gamma$, the value of $bc$ on the ground state in $(8.1)$.

\section{An Extension of the Knot Model}

The three models that have been described here (standard, knot, and preon) all exhibit the dynamics of the standard model but differ in their dependence on the SLq(2) representation assigned to the elementary fermion. In the standard, knot, and preon models, the elementary fermions are assigned to the $j=0$, $j=3/2$, and $j=1/2$ representations, respectively, of SLq(2) and these three models may in this sense be described as $j=0$, $3/2$, and $1/2$ realizations of SLq(2) by the standard model.

The empirical basis for the knot model depends on the natural way that the elementary $(t=1/2)$ fermions fit into the trefoil $(j=3/2)$ representation of the knot algebra. This property of the $j=3/2$ representation then suggests the examination of the $j=1$ and $j=1/2$ representations of the same algebra interpreted according to the same relations, namely:
\begin{equation}
\left( j, m, n \right) = 3 \left( t, -t_3, -t_0 \right)
\end{equation}
and
\begin{equation}
\left( j, m , n \right) = \frac{1}{2}\left( N, w, r+o \right)
\end{equation}
that fix D$_{mn}^j$ as in (2.7) and (2.2).

One then finds by (9.1) that there are four possible preons $(a,b,c,d)$, the four elements of D$_{mn}^{^{1/2}}$ where $a$ has charge $-e/3$ and $b$ is neutral, while $c$ and $d$ are anti-particles of $b$ and $a$ respectively. 

If (9.1) and (9.2) hold for all representations, one also finds that the structure of the sum (3.1) representing D$_{mm'}^j$ implies that every knotted particle, represented by D$_{mm'}^j$, may be regarded as a composite particle composed of $a,b,c,d$ preons, according to
\begin{equation}
\left( t, t_3, t_0, Q \right) = \sum_{p=(a,b,c,d)} n_p \left( t_p, t_{3p}, t_{0p}, Q_p \right)
\end{equation}
and
\begin{equation}
\left( N, w, \tilde{r} \right) = \sum_{p= (a,b,c,d)} n_p \left( N_p, w_p, \tilde{r}_p \right)
\end{equation}
for all representations, as in (3.9) and (4.6), where $n_p$ is the number of preons of type $p$.

At this point one may also ask whether it is necessary to explicitly postulate the correspondence $(9.2)$ with classical knots, since SLq(2) is already the knot algebra. Suppose we therefore drop this postulate to obtain a new model. This new model, which we term ``the extended knot model," is then \underline{totally determined by the knot algebra and the empirical}

\noindent \underline{rule (9.1) for interpreting this algebra}. While only the elementary particles of the standard model are described by the trefoil representation according to the knot model, \underline{no member of}

\noindent \underline{this representation or of any other representation is a priori excluded in the extended knot}

\noindent \underline{ model as a possible physical state}. In this alternate model we continue to assume that $\left( t, t_3, t_0 \right)$ and charge are determined by $(j,m,n)$ in (9.1) and that the number $\left( N' \right)$ of preons in each composite state is still $N'=6t=2j$ by $(3.4j)$ and $(3.5)$. For the complete trefoil representation one then finds, by (3.1), the Table 9.1 with the notation of (3.1) for A($s,t$).

\vspace{0.5cm}

\begin{center}
\bf{Table 9.1}
\end{center}
\begin{center}
{\underline{\bf{Elementary Fermions in the Extended Knot Model}}}
\end{center}
\begin{center}
\begin{tabular}{c|c|c|c|c}
\backslashbox[-20pt][l]{$m$}{$m'$} & $\frac{3}{2}$ & $\frac{1}{2}$ & $-\frac{1}{2}$ & $-\frac{3}{2}$ \\
 \hline
$\frac{3}{2}$ & $\begin{array}{c} \mbox{A}(3,0) \hspace{0.05cm} a^3 \\ \mbox{leptons} \end{array}$ & $\begin{array}{c} \mbox{A}(2,1) \hspace{0.05cm} a^2b \\ \overline{\mbox{up quarks}} \end{array}$ & $\begin{array}{c} \mbox{A}(1,0) \hspace{0.05cm} ab^2 \\ \mbox{down quarks} \end{array}$ & $\begin{array}{c} \mbox{A}(0,0) \hspace{0.05cm} b^3 \\ \overline{\mbox{neutrinos}} \end{array}$ \\
\hline
$\frac{1}{2}$ & $\begin{array}{c} \mbox{A}(2,1) \hspace{0.05cm} a^2c \\ \left( \hspace{0.05cm} \overline{\mbox{up quarks}} \hspace{0.05cm} \right)' \end{array}$ & $\begin{array}{c} \mbox{A}(2,0) \hspace{0.05cm} a^2d + \mbox{A}(1,1) \hspace{0.05cm} abc \\ \left( \mbox{down quarks} \right)' \end{array}$ &  $\begin{array}{c} \mbox{A}(0,1) \hspace{0.05cm} b^2c + \mbox{A}(1,0) \hspace{0.05cm} abd \\ \left( \hspace{0.05cm} \overline{\mbox{neutrinos}} \hspace{0.05cm} \right)' \end{array}$ & $\begin{array}{c} \mbox{A}(0,0) \hspace{0.05cm} b^2d \\ \overline{\mbox{down quarks}}\hspace{0.05cm}' \end{array}$ \\
\hline 
$-\frac{1}{2}$ & $\begin{array}{c} \mbox{A}(1,2) \hspace{0.05cm} \hspace{0.05cm} ac^2 \\ \left( \mbox{down quarks} \right)' \end{array}$ & $\begin{array}{c} \mbox{A}(0,2) \hspace{0.05cm} bc^2 + \mbox{A}(1,1) \hspace{0.05cm} acd \\ \left( \mbox{neutrinos} \right) ' \end{array}$ & $\begin{array}{c} \mbox{A}(1,0) \hspace{0.05cm} ad^2 + \mbox{A} (0,1) \hspace{0.05cm} bcd \\ \left( \hspace{0.05cm} \overline{ \mbox{down quarks}} \hspace{0.05cm} \right) ' \end{array}$ & $\begin{array}{c} \mbox{A} (0,0) \hspace{0.05cm} bd^2 \\ \left( \mbox{up quarks} \right)' \end{array}$ \\
\hline
$-\frac{3}{2}$ & $\begin{array}{c} \mbox{A}(0,3) \hspace{0.05cm} c^3 \\ \mbox{neutrinos} \end{array}$ & $\begin{array}{c} \mbox{A} (0,2) \hspace{0.05cm} c^2 d \\ \overline{\mbox{down quarks}} \end{array}$ & $\begin{array}{c} \mbox{A}(0,1) \hspace{0.05cm} cd^2 \\ \mbox{up quarks} \end{array}$ & $\begin{array}{c} \mbox{A}(0,0) \hspace{0.05cm} d^3 \\ \overline{\mbox{leptons}} \end{array}$
\end{tabular}
Charge $\sim (m + m^{\prime})$ $\hskip1.5cm$ Mass $\sim \overline{\mbox{D}}_{mm^{\prime}}^{\hskip0.07cm^{3/2}}\mbox{D}_{mm^{\prime}}^{^{3/2}}$
\end{center}

\vspace{0.5cm}
In Table (9.1) the overline means antiparticle and the prime means isotopic partner.

\noindent Only the first and fourth rows of Table 9.1 are occupied by the observed fermions and their antiparticles as described in the knot model. The second and third rows describe their isotopic partners. These isotopes are hypothetical three preon states with fractional charges (and their antiparticles) which are not particles of either the standard model or the knot model. These hypothetical particles, since they belong to the same representation as the quarks, may perhaps be explored at the same energies as the quarks. The complete D$_{mm'}^{^{3/2}}$ representation then contains a collection of preonic nuclei whose individual members are labelled by $(m,m')$, or equivalently by the charge $(-\frac{e}{3}(m+m'))$ and the mass operator $(\hspace{0.05cm} \overline{\mbox{D}}_{mm'}^{^{3/2}} \hspace{0.05cm} \mbox{D}_{mm'}^{^{3/2}} )$, as in the elements of the periodic table of nuclei, also labelled by mass and charge. The elements $(m, m^{\prime})$ and $(m^{\prime},m)$ are evidently isotopic.

The interpretation of these hypothetical particles depends on further assumptions about the preons. For example $a^2c$ and $a^2b$ have the same charge $(-\frac{2}{3}e)$ but different mass if the eigenvalues $\gamma$ and $\beta$ are different or if the $b$ and $c$ preons have different gluon charge.

%

%

The knot, the preon, and the extended preon model are all based on the knot algebra SLq(2) with physical interpretations according to (9.1). The physical interpretation of the knot and the preon model is further constrained by (9.2). In the alternative preon model, however, we drop this second constraint and therefore also drop this explicit correspondence with a classical knot. Hence the elementary particles of the knot model, i.e. leptons, neutrinos, and quarks, may be related to classical knots by (9.2) but this same correspondence does not hold for the alternative three particle states in Table 9.1.

Since there appear to be only 3 generations, the additional quark isotopes, if they exist, would be revealed in fine structure of the generational mass spectrum. Failure to find this fine structure may be interpreted as evidence for a correspondence between classical and quantum knots. Even in the extended model there are no leptonic isotopes beyond the three generations.

\section{The Preon Masses}
Eqn. (5.17) is an explicit expression for the masses of the four preons in terms of $q$, $\beta$, $\gamma$, and $\rho_p$ but this expression gives no indication of its magnitude. If the preon mass is very large, it must be compensated by a very large binding force, since the leptons, neutrinos, and quarks are relatively very light. One may perhaps attribute this strong binding to the $j=1$ vector preon field. In a different scenario as described by the rishon model of Harari and Seiberg$^{(6)}$ the preons are massless and bound by strong color and hypercolor forces. The nature of the field binding the preons is obviously a crucial question for preon models.

\section{Acknowledgements} 

I thank J. Smit, A.C. Cadavid, and J. Sonnenschein for helpful discussion.

%

\newpage
\vspace{2cm}

$\Large{\bf{References}}$
\begin{enumerate}
\item Finkelstein, R.J., An SLq(2) Extension of the Standard Model, arXiv:1205.1026v3 [hep-th]
\item Finkelstein, R.J., Solitonic Models Based on Quantum Groups and the Standard Model, arXiv:1011.2545v1 [hep-th]
\item H. Harari, Physics Letters B $\bf{\underline{86}}$ 83-86, M. Shupe, Physics Letters B $\bf{\underline{86}}$ 87-92
\item J. Smit, private communication
\item J. Sonnenschein, private communication
\item H. Harari and N. Seiberg, Nuclear Physics B204 (1982) 141-167
\end{enumerate}

\end{document}